\documentclass[11pt]{article}
\usepackage{moriond,epsfig,subfigure}

\def\be{\begin{equation}}
\def\ee{\end{equation}}
\def\bea{\begin{eqnarray}}
\def\eea{\end{eqnarray}}

\begin{document}
\vspace*{4cm}
\title{RECENT PROGRESS ON NNPDF FOR LHC}

\author{ M.~Ubiali$^1$, R.~D. Ball$^1$, L.~Del Debbio$^1$, S.~Forte$^2$, A.~Guffanti$^3$, J.~I.~Latorre$^4$, A.~Piccione$^2$, J.~Rojo$^5$~(NNPDF collaboration)}

\address{$^1$ Department of Physics, University of Edinburgh, Edinburgh, UK\\
$^2$ Dipartimento di Fisica, Universit\`a di Milano and INFN, Sezione di Milano, Milano, Italy\\
$^3$ Physikalisches Institut, Albert-Ludwigs-Universit\"at Freiburg, Freiburg, Germany \\
 $^4$ Departament d'Estructura i Constituents de la Mat\`eria, Universitat de Barcelona, Barcelona, Spain\\
 $^5$ LPTHE, Universit\'e Paris VI et Paris VII, Paris, France
%\\[10pt]
%\epsfig{file=me.eps,width=30mm}}
%\epsfig{file=maria2.eps,width=30mm}}
}
\maketitle
\vspace{-1.15cm}
\abstracts{ We present recent results of the NNPDF collaboration on a full DIS analysis of Parton Distribution Functions (PDFs). Our method is based on the idea of combining a Monte Carlo sampling of the probability measure in the space of PDFs with the use of neural networks as unbiased universal interpolating functions. The general structure of the project and the features of the fit are described and compared to those of the traditional approaches. }

\section{Introduction}

Experimental uncertainties in hadronic colliders are decreasing to a level where the most careful consideration has to be given to uncertainties in theoretical predictions. For hadronic processes involving high virtualities, the cross section can be written as a convolution of a partonic cross section and process-independent non perturbative functions, Parton Distribution Functions. The latter cannot be derived from perturbative QCD, however their evolution is. Consequently we can extract PDFs from one experiment and use them as a theoretical input for another. These are a key element for any phenomenological prediction and therefore a good knowledge of PDFs and of their errors is fundamental~\cite{heralhc}. 
 
Determining a set of functions with errors from a finite set of data points is far from being straightforward because of the many theoretical, experimental and phenomenological complications. The correct definition of an error band in the space of functions requires one to build up a probability density within that space. Let ${\cal F}[f_a(x)]$ be an observable depending on a PDF $ f_a(x)$; then its average value is defined as 
\begin{equation}
\label{eq:av}
\langle{\cal F}[f_a]\rangle = \int \,[{\cal D}f_a]\, {\cal P}[f_a(x)]\,{\cal F}[f_a(x)],
\end{equation}
where ${\cal P}$ is the probability density that one aims to determine. However this problem cannot be solved without introducing some theoretical assumptions, given that one is trying to infer an infinite amount of information from a finite number of points. The traditional solution consists in choosing a specific functional form such that the infinite-dimensional space of functions reduces to a finite-dimensional space of parameters which determine the chosen parametrisation.However the standard approaches~\cite{MRST,CTEQ,alekhin2} suffer of several drawbacks, mainly due to the lack of control
on the bias introduced the choice of a specific parametrisation and, more subtly, to the difficulty in providing a
consistent statistical interpretation of their PDFs uncertainties when dealing with non-gaussian uncertainties and incompatible data.

\section{NNPDF approach: main ingredients}
\label{sec:ing}

The difficulties mentioned in the previous section have stimulated various proposals for new approaches. The method proposed by the NNPDF collaboration was first tested on the parametrisation of DIS structure function data~\cite{f2ns,f2p}. Subsequently it was applied to the determination of a single non-singlet parton distribution~\cite{nnqns}. 
\begin{figure}[ht!]
\begin{center}
\includegraphics[width=6cm]{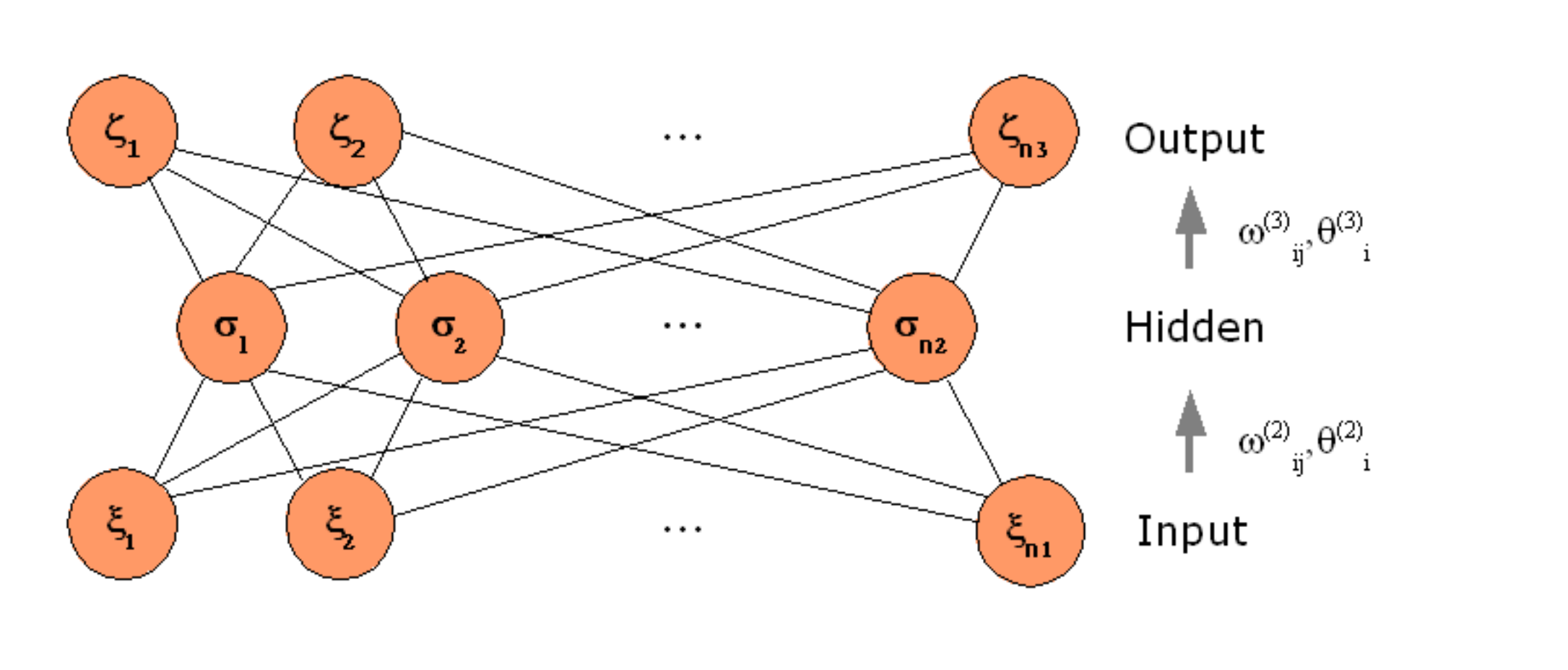}
\caption{Schematic representation of a multilayer feedforward neural network.} 
\label{fig:nn}
\end{center}
\end{figure}

Schematically one might divide the procedure typical of this approach into three main steps: first one generates $N_{\mathrm{rep}}$ Monte Carlo sets of artificial data about the original $N_{\mathrm{dat}}$ experimental points, distributed according to a multi-Gaussian distribution whose spread is determined by experimental errors. With ${\cal O}(1000)$ replicas the artificial sets reproduce central values, errors and correlations of the original data set within one per mil accuracy. Then for each replica one or more neural networks are trained in order to fit the artificial data; each neural network provides a parametrisation of one independent PDF at some fixed initial scale. The output of the neural network must be evolved to the scale of the experimental points in order to be compared to the data and in this way the parameters of the networks are fitted. Finally the set of $N_{\mathrm{rep}}$ trained neural networks provides a discrete representation of the probability density. Therefore Eq.~\ref{eq:av} reduces to an average over the ensemble of replicas,
\begin{equation}
\langle{\cal F}[f_a(x)]\rangle =\frac{1}{N_{\mathrm{rep}}}\sum_{k=1}^{N_{\mathrm{rep}}} {\cal F}[f_a^{(k)(\mathrm{net})}(x)].
\end{equation}
One ends up with a statistically accurate Monte Carlo determination of errors: the errors of PDFs or more generally of any observable depending on them is given by
 \begin{equation}
\sigma_{{\cal F}[f_a(x)]}=\sqrt{\langle{\cal F}[f_a(x)]^2\rangle-
\langle {\cal F}[f_a(x)]\rangle^2}; 
\end{equation}
in the same way one is able to evaluate any other statistical estimator.

It is important to outline that neural networks, Fig.~\ref{fig:nn}, provide nothing but a redundant and unbiased parametrisation for the PDFs at the initial scale: they are approximants whose functional form adapts well to a noisy and incomplete set of data.
Moreover, having constructed a Monte Carlo ensemble of replicas, it is easy to perform a variety of tests in order to assess the statistical consistency of our results as well as their independence on the details of the fitting procedure, {\it in primis} on the choice of parametrisation. The flexibility of neural networks architecture is particularly suitable to this kind of systematic analysis~\cite{nnqns}. 
\begin{figure}[ht!]
\begin{center}
\includegraphics[width=7cm]{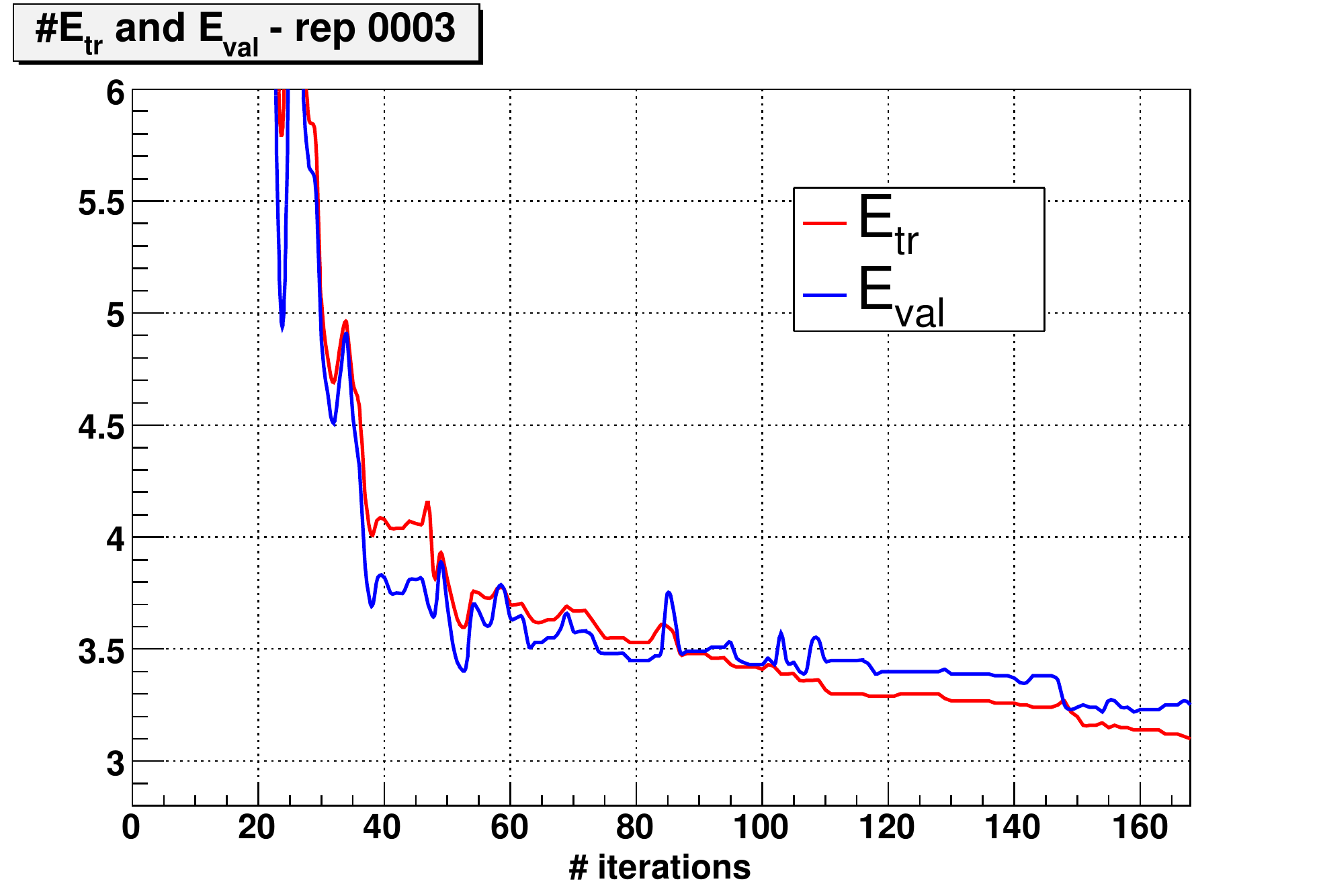}
\includegraphics[width=7cm]{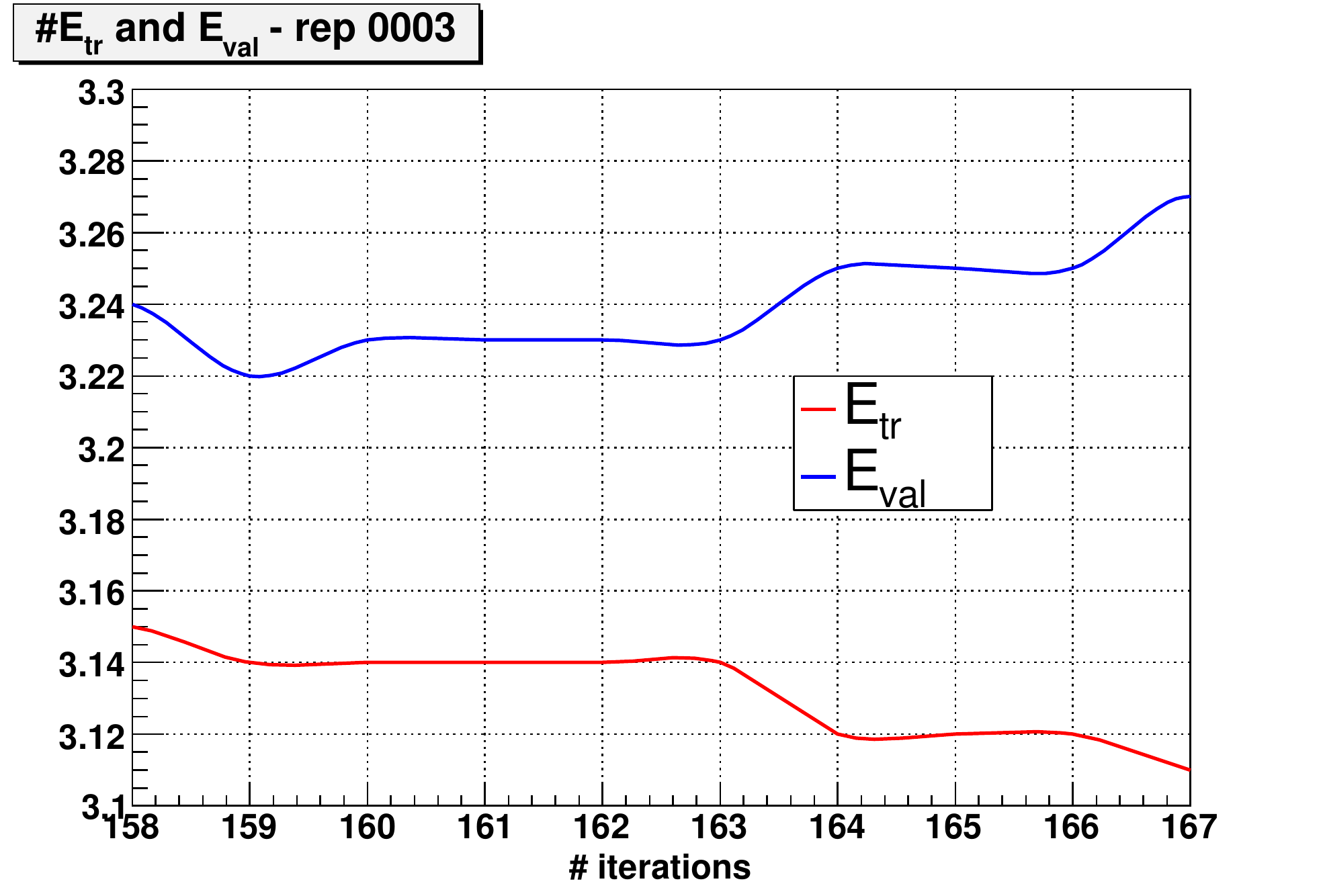}
\caption{Training and validation $\chi^2$ as a function of number of iterations for a given replica (left). Detail of the stopping region (right).}
\label{fig:ite}
\end{center}
\end{figure}
 In principle, any other redundant parametrisation with the same features would be suitable. 
 
 More specifically the redundancy is the essential feature of a parametrisation which ensures that one ends up with results minimally biased by the choice of the functional form. On the other hand the absolute minimum of the $\chi^2$ in a redundant parameter space would correspond to an over-learning of data, {\it i.e.} a regime where the parametrisation adapts not only to the physical behaviour but also to statistical fluctuations. Hence we need a criterion in order to stop the fit once having learnt the physical behaviour but before over-learning the data. To do this, for each replica we divide randomly the data into two sets: training and validation. The first set is the one on which we actually perform the fit by minimising the fully correlated $\chi^2$; the latter is also evaluated over the validation set at each iteration of the minimisation. When the training $\chi^2$ is still decreasing and the validation one starts increasing, as it is shown in Fig.~\ref{fig:ite}, we are entering in the over-learning region and therefore the fit must be stopped. 

\section{DIS analysis}

The extension of the non singlet fit~\cite{nnqns} to a full DIS analysis is the upcoming result of the NNPDF
collaboration.  
\begin{figure}[ht!]
\begin{center}
\includegraphics[width=9cm]{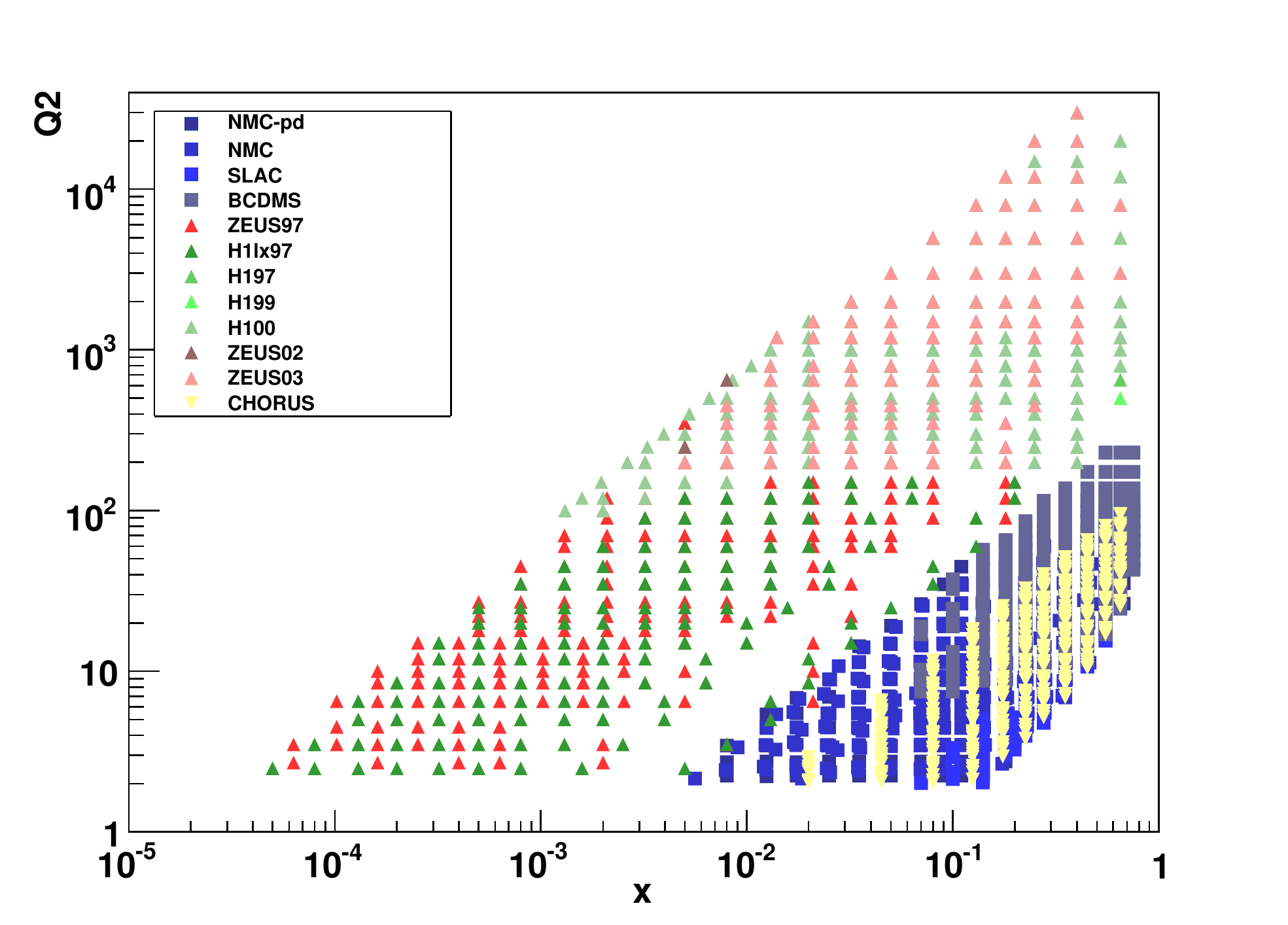}
\caption{Kinematic range of experiments included in the fit.}
\label{fig:kin}
\end{center}
\end{figure}
This extension involves the significant complication of going from one PDF and one observable to a set of parton
distributions extracted from many different experiments and observables. DIS and neutrino data, structure functions and reduced cross sections are included, as shown in Fig.~\ref{fig:kin}. 
For this first fit we have used a smaller set of PDFs by imposing some flavor assumptions like the symmetric strange sea $s(x)=\bar{s}(x)$ and the proportionality between strange and non-strange seas $\bar{s}(x)=\frac{C}{2} (\bar{u}(x)+\bar{d}(x))$ \footnote{ $C\sim 0.5$ as suggested by dimuon data}. With these assumptions we parametrise four combinations of quarks and the gluon distribution at a fixed initial scale ($Q_0^2=2$ GeV$^2$) by mean of five multilayer neural networks: $
\Sigma(x)$, $V(x)\equiv (u_v + d_v)(x)$, $T_3(x)\equiv  (u+\bar{u} -d-\bar{d})(x)$, $\Delta_S(x)\equiv ( \bar{d} -\bar{u})(x)$, $g(x)$. The overall normalisation of the nets is fixed by imposing momentum and valence sum rules.
\begin{figure}[ht!]
\begin{center}
\includegraphics[width=7cm]{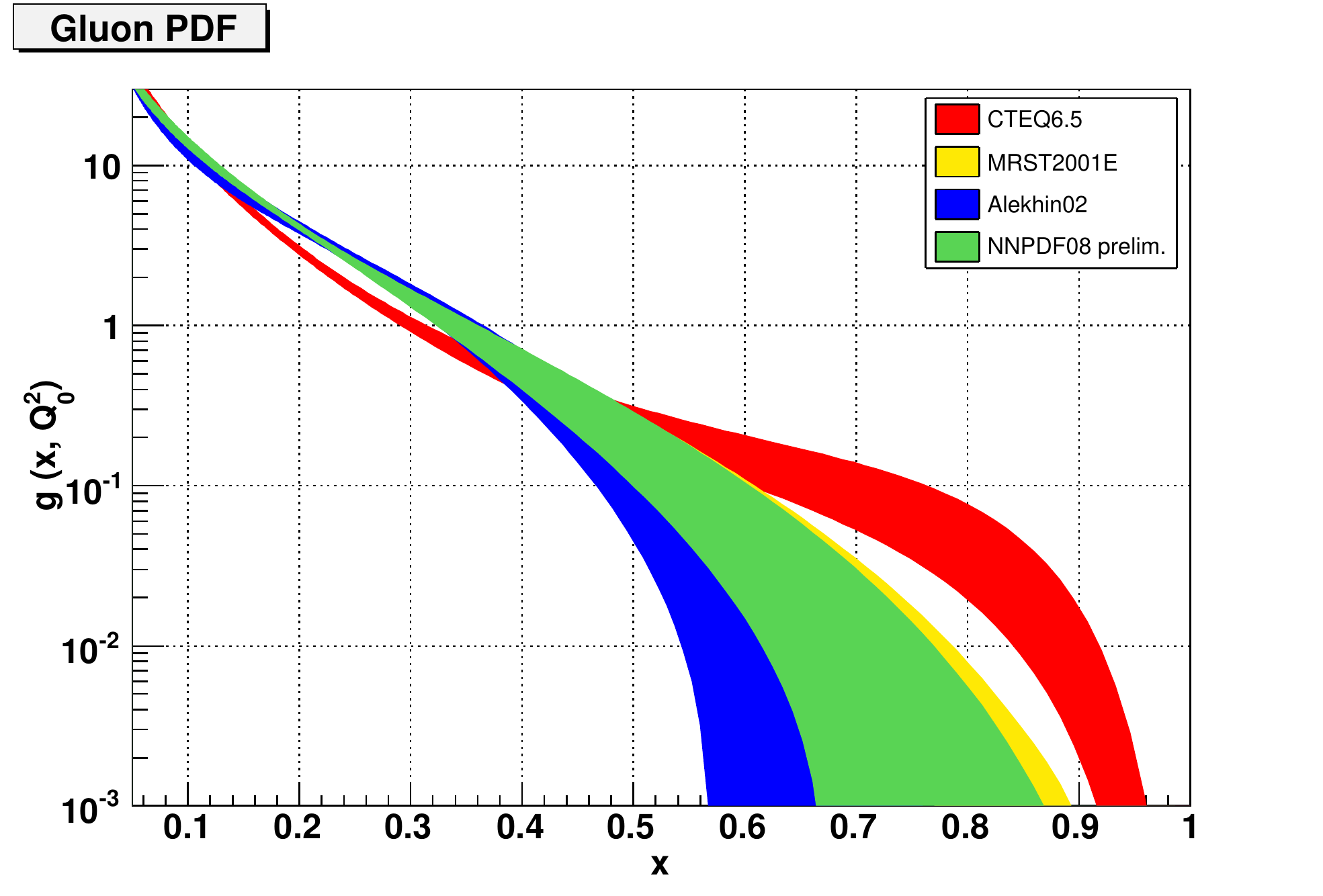}
\includegraphics[width=7cm]{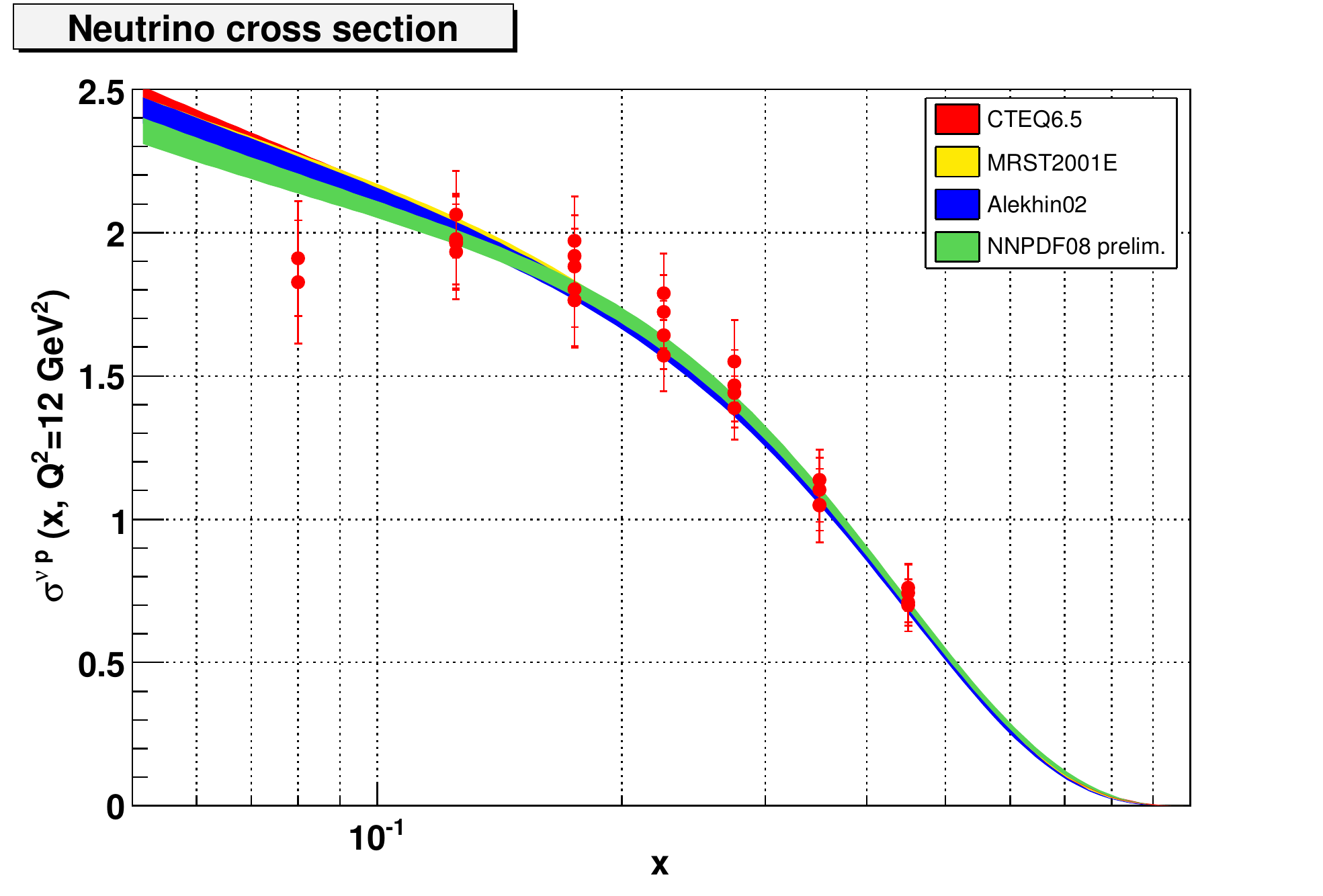}
\caption{Gluon at $Q^2_0=2$GeV$^2$(left). Reduced neutrino cross section at $Q^2=12$GeV$^2$(right). Preliminary results.}
\label{fig:plot}
\end{center}
\end{figure}
In the present code we perform a hybrid N-space and x-space Next-to-Leading Order evolution; the solution of the DGLAP equations is calculated analytically into the Mellin space, using a Zero-Mass Variable Flavour Number scheme and including Target Mass Corrections. Then it is inverted back into the x space and convoluted with the initial x-space parton distributions provided by the nets. 

From our preliminary results, Fig.~\ref{fig:plot}, we see that while in the data region the different approaches are compatible, in the extrapolation region they do deviate from each other. Interestingly our fit produces results consistent with those obtained by the other collaborations~\cite{MRST,CTEQ,alekhin2} and our error bands tend to get bigger in the region where data do not constrain PDFs behaviour. These results suggest that standard approaches to PDFs fitting might lead to an underestimation of errors associated with parton densities and that our combination of MC techniques and neural networks is a feasible alternative as well as adequate for detailed statistical studies. Having published the full DIS fit, the next step will be the inclusion of more data sets and some phenomenological studies on the impact of our error bands on the main LHC observables.

\section*{Acknowledgments} 

M.~Ubiali is funded by a SUPA studentship.

\end{document}